# Canonical Quantization, Quasi-Hermiticity, Observables and the Construction of Mutually Unbiased Bases and Other Complete Sets

Donald J. Kouri
Department of Physics and Department of Mathematics
University of Houston
Cullen Blvd.
Houston, TX 77204-5005

Cameron L. Williams
Department of Mathematics
University of Houston
Cullen Blvd.
Houston, TX 77204-5003

Nikhil Pandya
Department of Physics and Department of Mathematics
University of Houston
Cullen Blvd.
Houston, TX 77204-500

## ABSTRACT

Using ideas based on supersymmetric quantum mechanics, we design canonical transformations of the usual position and momentum to generate generalized "Cartesian-like positions, $W$, and momenta, $p_W$" with unit Poisson brackets. These are quantized by the usual replacement of the classical $x$, $p_x$ by quantum operators, leading to an infinite family of potential "operator observables". The fundamental issue is that all but one of the resulting operators are not Hermitian (formally self-adjoint) in the original position representation. By using Dirac quantization, we show that the resulting operators are "quasi-Hermitian" relative to the $x$-representation and that all are Hermitian in the $W$-representation. Depending on how one treats the Jacobian of the canonical transformation in the expression for the classical momentum, $p_W$, quantization yields a) continuous mutually unbiased bases (MUB) b) orthogonal bases (with Dirac delta normalization) c) biorthogonal bases (with Dirac delta normalization) d) new $W$-harmonic oscillators yielding standard orthonormal bases (as functions of $W$) and associated coherent states and Wigner distributions. The MUB include $W$-generalized Fourier transform kernels whose eigenvectors are the $W$-harmonic oscillator eigenstates, with the spectrum $(\pm 1, \pm i)$, as well as "W-linear chirps". The $W$, $p_W$ satisfy the uncertainty product relation: $\Delta W \Delta p_W \geq 1/2$, $\hbar = 1$.

## I. Introduction

Basis sets are of prime importance in quantum mechanics because they are fundamentally connected to observables. The observables serve to dictate the relevant operators in quantum mechanics, which are typically required to be self-adjoint. Some reasons underlying this are: (1) a basic postulate of quantum mechanics is that the results of measurements must be real and the eigenvalues of the relevant operator are the only possible results when measuring an observable (2) the eigenstates of observables must be complete, spanning the relevant physical Hilbert space of possible states (as a result of the requirement that their eigenvalues are the only possible results of measurements) (3) these eigenstates can always be arranged to be mutually orthogonal (because measurements lead to an orthogonal projection of the state of the system onto an eigenstate of the observable's operator). The requirement that quantum mechanical operators for observables be self-adjoint is sufficient to ensure that the above properties hold, but one can ask "is it also necessary"? Certainly, an examination of the textbooks and most of the literature dealing with quantization of classical mechanics would suggest that it is [1-7]. However, interesting results have been obtained recently using non-self adjoint Hamiltonians (besides those describing meta-stable systems) [8-20]. In the course of these studies, attention has focused on the Hamiltonian operator and systems with discrete spectra. In this paper, we begin exploring systematic quantizations that lead to non-self adjoint, but "quasi-Hermitian" operators [8-20] that potentially may describe observables similar to generalized "position" and "momentum".

In addition, complete sets of functions are also of great importance for the computational aspects of quantum mechanics. For most systems of physical interest, exact solutions are not possible and therefore approximations are the "order of the day". A great many of these rely on finding the best possible basis set to use in expanding the desired solutions (e.g., both in perturbation theory and variational calculations). It follows that one is always interested in finding new, more optimal basis sets for applications. This connection between observables and complete bases in quantum mechanics suggests that one should also seek the optimum variables (observables) to describe the system of interest.

Yet another area where basis sets play an important role is in quantum optics. In this case, coherent states (which typically are over-complete, non-orthogonal bases) are the center of focus. Especially of interest are those associated with operators that form a Lie algebra (for constructing displacement operators) or that are eigenstates of (non-self adjoint) annihilation operators.

Most recently, the field of quantum computing and quantum cryptography has generated interest in finding new groups of basis sets that are "mutually unbiased" [21-24]. Although most efforts have concentrated on finite dimensional bases, there is also great interest in discovering continuously infinite systems, the prime example of which are the (improper) eigenstates of the three self adjoint operators $\hat{x}$, $\hat{p}_x$, $\hat{x} + \hat{p}_x$. A fundamental property of $\hat{x}$ and $\hat{p}_x$ is that the minimum uncertainty product $\Delta x \Delta p_x$ is generated by the Gaussian. This result is due to the fact that the minimizing state, $|\psi_{min}\rangle$, satisfies $\hat{x} |\psi_{min}\rangle \propto \hat{p}_x |\psi_{min}\rangle$. The MUB nature of the eigenstates of the set $\hat{x}$, $\hat{p}_x$, $\hat{x} + \hat{p}_x$ appears to be related to the fact that exact knowledge of $x$ ($p_x$) implies infinite uncertainty in $p_x$ ($x$). This is related to the fact that $[\hat{x}, \hat{p}_x] = i \hat{1}$. One suspects that this must be a characteristic required to generate continuous MUB [23]. Indeed, there is a number of important types of complete basis sets associated with position and momentum and they are relevant in various areas of interest. The position and momentum representations are ubiquitous in quantum mechanics. In addition, the wave function $\langle x | p_x \rangle = \dfrac{e^{ikx}}{\sqrt{2\pi}}$ is fundamental not only in quantum mechanics but also in all the sciences, engineering, signal processing, probability theory and diffusion processes, medicine, etc. In addition, the harmonic oscillator (with its symmetric quadratic dependence on the position and momentum operators) provides a complete set of eigenstates. These also are eigenfunctions of the Fourier transform and the ground state minimizes the product of the position and momentum uncertainties. This uncertainty product minimizing state leads to one particular realization of coherent states (there are many others, including an infinite variety arising from application of a displacement operator (in terms of the position and momentum operators) to any normalizable "fiducial" state [33]).

In light of the key role played by the position and momentum coordinates in obtaining complete bases of various types, it is natural to inquire as to the role that is played by canonical transformations of the type $x \to W$, $p_x \to p_W$, such that the Poisson bracket satisfies $\{W, p_W\} = 1$. This will also naturally involve how one proceeds from classical to quantum mechanics (we shall use the Dirac canonical quantization procedure). We will explore how canonical transformations can be used to create new and unusual complete bases, which may be orthogonal, biorthogonal or MUB, as well as new coherent states, Wigner distribution functions and generalized Fourier transform kernels.

## II. Classical Dynamical Considerations
Our approach to the construction of new, complete sets of basis functions has its foundation in the concept of canonically conjugate variables and canonical transformations. In this study, we shall focus on point transformations. As usual, we begin with the standard canonically conjugate variables, $x$ and $p_x$. These variables are characterized by the property that their Poisson bracket equals one. Dirac based an approach to quantization in which the canonical variables are replaced by appropriate operators, the Poisson bracket is replaced by the commutator of the relevant operators and the scalar "1" is replaced by $i\hbar$ times the identity operator [1,7]. As is well known, the quantum mechanical operators $\hat{x}$, $\hat{p}_x$ provide the standard, continuous complete sets of orthonormal (under the Dirac delta normalization) "eigenvectors". Adding the eigenstates of the combined $\hat{x} + \hat{p}_x$ operator, these complete basis sets constitute a group of mutually unbiased bases (MUB) [21-24]. It appears that a key component needed for this property is that the position and momentum are canonically conjugate classical variables, which when quantized, yield a constant quantum mechanical commutator. We speculate that this property is a necessary (but not sufficient) condition to obtain complete sets of eigenstates that are MUBs [23]. There is, of course, a long history of canonical transformations in classical dynamics as the means of constructing "natural" generalized, canonically conjugate coordinates in terms of which the dynamics is the simplest [25-29]. Many examples exist where the natural dynamical variables are quite different from the original $x$ and $p_x$, and their discovery generally reflects some fundamental feature of the dynamics. In this paper, we shall explore replacing the Cartesian position and momentum by new, canonically conjugate variables and then, using Dirac quantization, explore the properties of the resulting quantum mechanical operators and their (improper) eigenstates. In this process, we shall explore both manifestly self adjoint and apparently non-self adjoint operators.

To do this, we require some guide as to possible reasonable choices for new "displacement or position variables" and their canonically conjugate momenta (once the "position-like" variable is decided upon, the conjugate momentum is determined by requiring its Poisson bracket with the new "position" be equal to 1). A hint for choosing such new variables can be found in super symmetric quantum mechanics (SUSY) [30-32]. In one dimensional SUSY, the nodeless ground state for a system (in the domain $-\infty < x < \infty$) is expressed as

$$\psi_0(x) = \psi_0(0) \exp[-\int_0^x dx' W(x')] , \quad (1)$$

which has a SUSY ground state energy equal to zero. As discussed in earlier work [31], this state minimizes the uncertainty product $\Delta W \Delta p_x$. However, the commutator of W with the standard momentum operator is proportional to $\frac{dW}{dx}$, which is, in general, constant only for $W = x$ (plus an arbitrary constant that simply shifts the energy levels). Thus, in general, $p_x$ and $W$ are not canonically conjugate variables. Equation (1) also implies the relation

$$V = W^2 - \frac{dW}{dx} = W^2 + \frac{i}{\hbar}\left[\hat{W}, \hat{p}_x\right] \quad (2)$$

between the physical potential and W. It is the fact that $\hat{p}_x$ and $\hat{w}$ are not canonically conjugate operators that is responsible for the "anharmonicity" of the system. However, this relation of W to the potential energy has led the SUSY literature to designate W as a "super potential" [30]. Recent work [31] has shown that an alternate but profitable way to interpret W is as a "generalized position" or "displacement variable". This is reinforced by the observation that the quintessential example for W is that for the harmonic oscillator:

$$W(x) = x \quad . \quad (3)$$

Clearly, the general W can be considered a position variable and not a potential, super or otherwise. This view has led to the development of new "system adapted Klauder-Skagerstam coherent states" [31,33]. These new coherent states have been shown to provide a superior basis set (compared to standard Gaussian-based coherent states or to a standard harmonic oscillator basis) for computing accurate excited state energies for some polynomial choices of W (which result in anharmonic oscillators). More recently, Williams, et. al. [34-35] have shown that completely new one dimensional, generalized harmonic oscillator systems, characterized by the su(1,1) Lie algebra, can be constructed. It is interesting to note that this can be based on interpreting W as a generalized position, arising from a canonical transformation.

This suggests that a useful family of W's might be polynomials in x, whose lowest and highest powers are odd, and with non-negative coefficients (to ensure monotonicity and that the domain remains $(-\infty, \infty)$). We shall restrict ourselves to W's of the form

$$W(x) = \sum_{j=1}^{2J+1} a_j x^j \, , \, a_j \geq 0 \quad . \quad (4)$$

The coefficients are also restricted such that the coefficient of each even power is never larger than the coefficient of the next lower odd power. This ensures that the ground state is not only normalizable but that the $x \to W$ transformation is one-to-one and onto (it has a derivative that is non-negative so it is a monotonic function and invertible). The original (classical) canonical variables are $(x, p_x)$. We then require that the new, canonically conjugate momentum, $p_W$, be a function of $(x, p_x)$ such that the Poisson bracket satisfies

$$\{W, p_W\} = 1 \quad . \quad (5)$$

**This is easily solved to yield the general classical expression**

$$p_W \equiv \frac{1}{\left(\frac{dW}{dx}\right)^{1-\alpha}} p_x \frac{1}{\left(\frac{dW}{dx}\right)^{\alpha}} + g(x) \, , \quad (6)$$

where $g(x)$ is the constant of integration of Eq. (5) along a path of constant x. We invoke the simplest assumption of taking the solution with $g(x) = 0$. Thus, the Jacobian of the transformation can be split in infinitely many ways classically since all the quantities commute. (The condition Eq. (1) remains true for any choice of α.) It is immediately clear, of course, that this expression suffers from the usual issue that when quantized, it will typically result in non-self adjoint momentum operators in the x-representation, with measure "$dx$". Indeed, a common procedure is to define the new canonical momentum as the arithmetic average

$$p_W = \frac{1}{2}\left[\frac{1}{\left(\frac{dW}{dx}\right)^{1-\alpha}} p_x \frac{1}{\left(\frac{dW}{dx}\right)^{\alpha}} + \frac{1}{\left(\frac{dW}{dx}\right)^{\alpha}} p_x \frac{1}{\left(\frac{dW}{dx}\right)^{1-\alpha}}\right]. \quad (7)$$

Clearly, this leads to an x-representation, self-adjoint operator when quantized. We also stress that Eq. (6) alone yields a manifestly self-adjoint momentum operator for the choice $\alpha = 1/2$, which is also included in Eq. (7). **It is important to note, however, that all of these definitions of the classical canonical momentum yield a Poisson bracket equal to 1.** We now have the classical canonically conjugate pair $\{x, p_x\}$ and an infinity of possible new, canonically conjugate variables $\{W(x), p_W\}$. Due to the $\frac{dW}{dx}$ factors, it is most convenient to quantize in the x-representation and choosing to quantize with Eq. (6) or (7) will result in different quantum operators, whose (improper) eigenstates are a subject of our investigations.

### III. Dirac Quantization of the Canonically Conjugate Classical Variables

We now turn to consider the quantization of the new canonically conjugate variables. But we are immediately faced with the usual ambiguity that there are infinitely many equally valid classical expressions for the new canonical momentum. To illustrate, consider quantizing the choices $\alpha = 0$ *and* 1 for Eq. (6) by replacing the classical variables by the standard quantum mechanical position and momentum operators:
$\alpha = 0$:

$$\hat{p}_W = \frac{-i}{dW/dx} \frac{d}{dx}, \quad (8)$$

and $\alpha = 1$:

$$\hat{p}_W = -i \frac{d}{dx} \frac{1}{dW/dx} \quad . \quad (9)$$

**These are clearly not self-adjoint operators in the x-representation. (Indeed, Eq. (9) is the adjoint of Eq. (8).) Therefore, one would normally discard both.** However, we note that for **all** of the infinitely many definitions of the canonical momenta (including Eq. (7)), their Poisson bracket with W is equal to 1 (see Eq. (5)). **Thus, Eqs. (6) and (7) yield the Dirac quantization result**

$$\left[ \hat{W}, \hat{p}_W \right] = i \hat{1} \quad . \quad (10)$$

**Appealing to Occam's razor, this immediately implies that the corresponding quantum operators in the W-representation are**

$$\hat{W} = W \quad , \quad \hat{p}_W = -i \frac{d}{dW} \quad . \quad (11)$$

**Interestingly, employing the chain rule for derivatives, Eq. (8) also directly yields** $\frac{-i}{dW/dx} \frac{d}{dx} \equiv -i \frac{d}{dW}$. So $\hat{W}, \hat{p}_W$ are obviously self-adjoint operators in the W-representation! In fact, $\hat{p}_W$ is a quasi-Hermitian operator [8-20] relative to the original x-representation momentum operator, i.e., $\hat{p}_x = \frac{dW}{dx} \hat{p}_W$ or $dx \, \hat{p}_x = dW \, \hat{p}_W$.

Furthermore, $\frac{dW}{dx}$ is a non-negative polynomial and serves as a metric for the measure $dW = dx \frac{dW}{dx}$. We stress that the measure and metric are supplied automatically by the canonical transformation. Furthermore, the structure of Eq. (11) is identical to the original structure for the operators $\hat{x}, \hat{p}_x$ [23]. It is then clear that there exist complete, orthonormal (in the Dirac delta function sense) eigenstates of the three operators $\hat{W}, \hat{p}_W, \hat{W} + \hat{p}_W$ which will constitute MUB! Their structure is given by

$$\langle W'|W\rangle = \delta(W'-W) = \frac{dW}{dx}\delta(W(x)-W(x')), \quad (12)$$

$$\langle W|p_W\rangle = \frac{e^{ip_W W}}{\sqrt{2\pi}} = \frac{e^{iW(k)W(x)}}{\sqrt{2\pi}}, \quad (13)$$

$$\langle W|W+p_W\rangle = \frac{e^{ip_W W - iW^2/2}}{\sqrt{2\pi}} = \frac{e^{iW(k)W(x) - iW(x)^2/2}}{\sqrt{2\pi}}, \quad (14)$$

which are clearly MUB [23]. (The second parts of Eqs. (12) – (14) are in the x-representation and the second parts of Eqs. (13) - (14) involve $W(k) = p_W$ because the argument of the exponential, $p_W W$ must be dimensionless and we require that the two eigenvalues have the same structure.) Equation (14) is recognized as a linear chirp. We stress that when using the above for quantum cryptography, one should work in the original x-coordinate representation since this adds an additional layer of security because there are infinitely many choices for W and each one results in a different x-representation MUB. Of course, the measure is automatically taken into account through the chain rule expression $dW = dx\frac{dW}{dx}$. In addition, we note that the minimum uncertainty product condition, $\Delta x \Delta p_x = 1/2$, is now generalized to $\Delta W \Delta p_W = \Delta W(x) \Delta W(k) = 1/2$! **Thus, we arrive at the result that even though all the choices, Eq. (6), for the canonically conjugate momentum (except for $\alpha = 1/2$) lead to non-self adjoint (quasi-Hermitian) momentum operators in the coordinate representation, under the assumption of greatest simplicity and by direct application of Dirac quantization to the Poisson bracket, they all result in the same self-adjoint "position" and "momentum" operators in the W-representation. This is also true for the self-adjoint choices, Eq. (7),** *and therefore all choices lead to an unique MUB for each possible W***!** We shall now explore the interesting issue of self-adjointness, and its lack, in the coordinate representation of the new operators.

### IV. Lack of Self-Adjoint Property for Canonical Variables Not Quantized by Dirac's Method

We shall quantize the infinite set of canonically conjugate variables that result in non-self adjoint operators in the coordinate representation (along with those that do result in self-adjoint quantum operators, but, as we shall see, do not directly yield MUB in the coordinate representation). Interestingly, it is easily shown that the commutator of $\hat{w}$ and **any** of the possible coordinate representation "momentum operators" equals $i\hbar\hat{1}$.

For example, as Eq. (8) stands, it clearly results in a momentum operator which is not self-adjoint in the x-representation. However, as discussed above, based on the "chain rule" for derivatives, it has the precise, self-adjoint form

$$\hat{p}_W = -i \frac{d}{dW} \quad (15)$$

in the W-representation. Thus, although Eq. (8) is not self-adjoint in the x-representation, it is in the W-representation. In fact, this is true for all the possible choices of $p_W$, provided that one assumes that the only sensible quantization of $W$ in the W-representation is $\hat{W} = W$ and this, along with Eq. (10) and Occam's razor, implies Eq. (15). This is evidently self-adjoint, but only with respect to the measure $dW$. **This suggests that Eq. (8) is a valid way to quantize systems such that one obtains a self-adjoint momentum operator without symmetrizing (albeit with a new measure)!** It is instructive to consider the self-adjoint property of Eq. (11) in more detail. It is clear that

$$\int_{-\infty}^{\infty} dW \, \psi^*(W) \left[-i \frac{d}{dW}\right] \phi(W) = \int_{-\infty}^{\infty} dW \, \phi(W) \left[-i \frac{d}{dW}\right] \psi^*(W) \quad . \quad (16)$$

However, in the x-representation, this is exactly equal to

$$\int_{-\infty}^{\infty} dx \frac{dW}{dx} \tilde{\psi}^*(x) \left[\frac{-i}{dW/dx} \frac{d}{dx}\right] \tilde{\phi}(x) = \int_{-\infty}^{\infty} dx \, \tilde{\psi}^*(x) \left[-i \frac{d}{dx}\right] \tilde{\phi}(x) \quad (17)$$

$$= \int_{-\infty}^{\infty} dx \, \tilde{\phi}(x) \left[-i \frac{d}{dx}\right] \tilde{\psi}^*(x) \quad (18)$$

$$\equiv \int_{-\infty}^{\infty} dx \frac{dW}{dx} \tilde{\phi}(x) \left[\frac{-i}{dW/dx} \frac{d}{dx}\right] \tilde{\psi}^*(x) \quad . \quad (19)$$

Thus, we see explicitly that Eq. (8) is self-adjoint under a measure which we again stress is automatically dictated by the canonical transformation. This is extremely interesting and suggests that we should explore the x-representation adjoints of other choices for the canonical momentum. To do this, we treat the general expression, Eq. (6) since it includes all the obvious ways to split the Jacobian factor. Of course, the symmetrized expression, Eq. (7) is obviously self-adjoint in the coordinate representation. We consider the integral

$$\int_{-\infty}^{\infty} dx\, \phi^*(x)\, \frac{-i}{\left(\dfrac{dW}{dx}\right)^{1-\alpha}}\, \frac{d}{dx}\, \frac{1}{\left(\dfrac{dW}{dx}\right)^{\alpha}}\, \psi(x) \quad . \quad (20)$$

We define

$$\bar{\phi}^* = \frac{\phi^*}{\left(\dfrac{dW}{dx}\right)^{1-\alpha}}, \quad (21)$$

$$\bar{\psi} = \frac{\psi}{\left(\dfrac{dW}{dx}\right)^{\alpha}}, \quad (22)$$

and it immediately follows that

$$\int_{-\infty}^{\infty} dx\, \psi(x)\, \frac{-i}{\left(\dfrac{dW}{dx}\right)^{\alpha}}\, \frac{d}{dx}\, \frac{1}{\left(\dfrac{dW}{dx}\right)^{1-\alpha}}\, \phi^*(x) \quad . \quad (23)$$

It is obvious that unless $\alpha = 1/2$, the momentum operator is not self-adjoint in the x-representation. This apparently unsatisfactory situation is fundamentally related to the role of the measure, which arises due to the Jacobian of the canonical transformation.

Our next consideration is to determine whether the operators in Eqs. (6) - (7), are self-adjoint under the measure $dx\, \dfrac{dW}{dx}$. We only need to treat Eq. (6) directly since the results will dictate those for Eq. (7). We thus consider the integral

$$I_\alpha = \int_{-\infty}^{\infty} dx \left(\frac{dW}{dx}\right) \psi^*(x) \frac{1}{\left(\frac{dW}{dx}\right)^{1-\alpha}} \hat{p}_x \frac{1}{\left(\frac{dW}{dx}\right)^{\alpha}} \phi(x) \quad (24)$$

$$= \int_{-\infty}^{\infty} dx\, \psi^*(x) \frac{1}{\left(\frac{dW}{dx}\right)^{-\alpha}} \hat{p}_x \frac{1}{\left(\frac{dW}{dx}\right)^{\alpha}} \phi(x) \quad (25)$$

$$= \int_{-\infty}^{\infty} dx\, \phi(x) \frac{1}{\left(\frac{dW}{dx}\right)^{\alpha}} \hat{p}_x \frac{1}{\left(\frac{dW}{dx}\right)^{-\alpha}} \psi^*(x) \quad (26)$$

$$= \int_{-\infty}^{\infty} dx \left(\frac{dW}{dx}\right) \phi(x) \frac{1}{\left(\frac{dW}{dx}\right)^{1+\alpha}} \hat{p}_x \frac{1}{\left(\frac{dW}{dx}\right)^{-\alpha}} \psi^*(x) \quad (27)$$

This shows that unless $\alpha = 0$, $\hat{p}_W$ in Eq. (6) is not self-adjoint under the measure $dW = dx\left(\frac{dW}{dx}\right)$. We arrive at the results that:

(1) For Eq. (6) with $\alpha = 1/2$ and Eq. (7), $\hat{p}_W$ is self-adjoint under the measure $dx$ (i.e., in the x-representation)
(2) For $\alpha \neq 1/2$, Eq. (6) is not self-adjoint under the measure $dx$ (i.e., in the x-representation)
(3) For $\alpha = 0$, Eq. (6) is self-adjoint under the measure $dW$ (i.e., in the W-representation)
(4) For $\alpha \neq 0$, Eq. (6) (and also clearly for Eq. (7)) is not self-adjoint in the W-domain
(5) For $\alpha = 0$, Eq. (6) gives rise to continuous MUB for the operators
$$\hat{W},\; -i\frac{d}{dW},\; \hat{W},\; -i\frac{d}{dW}$$

We now turn to show explicitly that, in fact, the role of the measure is to ensure that these non-self adjoint (as well as the self-adjoint) choices of momentum operators give rise to complete, continuous x-representation, biorthogonal (or orthogonal in the self-adjoint cases) basis sets with Dirac delta normalization.

**V. Continuous Basis Sets for Canonically Conjugate, Non-Self Adjoint Operators in the x-Representation**

The key to understanding the role of the Jacobian and measure for the improper eigenstates of the new, non-self adjoint momentum operators is to evaluate them explicitly in the coordinate representation. This is easily done for the general case of Eq. (6). Thus, we solve the eigenequation

$$\left[ \frac{-i}{\left(\frac{dW}{dx}\right)^{1-\alpha}} \frac{d}{dx} \frac{1}{\left(\frac{dW}{dx}\right)^{\alpha}} \right] \phi_{p_W}(x) = p_W \, \phi_{p_W}(x) \quad (28)$$

where $p_W$ is a constant (eigenvalue). We recognize that, in general, the above momentum operator ($\alpha \neq 1/2$) is not self-adjoint in the x-representation. However, it turns out that even when one allows complex eigenvalues, $p_W$, the new states are not normalizable, even in the Dirac delta sense. As a result, we need only consider real eigenvalues for Eq. (28). It is easily shown that the solution is

$$\langle x | \phi_{p_W} \rangle = \phi_{p_W}(x) = \frac{\left(\frac{dW}{dx}\right)^{\alpha}}{\sqrt{2\pi}} e^{iW(k)W(x)} \quad . \quad (29)$$

Here, we again replace the eigenvalue $p_W$ by $W(k)$ since the argument of the exponential must be dimensionless. In similar fashion, we also easily find that the dual eigenvector is

$$\langle \phi_{p_W} | x \rangle = \phi^*_{p_W}(x) = \frac{\left(\frac{dW}{dx}\right)^{1-\alpha}}{\sqrt{2\pi}} e^{-iW(k)W(x)} \quad . \quad (30)$$

We then see that the eigenstates satisfy the completeness relation

$$\langle \phi_{\bar{p}_W} | \phi_{p_W} \rangle = \frac{1}{2\pi} \int_{-\infty}^{\infty} dx \left(\frac{dW}{dx}\right)^{1-\alpha+\alpha} e^{i(p_W - \bar{p}_W)W(x)} = \delta(\bar{p}_W - p_W) \quad . \quad (31)$$

Clearly, the eigenstates and their duals **automatically** supply the required factors to produce the correct measure, leading to a **complete biorthogonal (or orthogonal when $\alpha = 1/2$ ) basis**.

This is analogous to the situation studied by Kouri, et. al [15] where they showed that the harmonic oscillator Hamiltonian could be transformed similarly to a non-self adjoint form, leading to eigenstates that belonged to a biorthogonal complete set. However, it is also clear that these bases will not be part of any MUB because there is always a non-constant-modular factor in the eigenvector and/or its dual. Furthermore, even for the self-adjoint choice $\alpha = 1/2$, there will not be an MUB. In particular, we also see that neither of the choices $\alpha = 0$ or $\alpha = 1$ leads to an MUB in the coordinate representation. It follows from this that the same is true for the other self-adjoint choice, Eq. (7).

It is also interesting to determine the x-representation eigenstates of Eq. (7) (which includes Eq. (6) for $\alpha = 1/2$) for the general, self-adjoint $\hat{p}_W$. We consider

$$\frac{1}{2}\left[\frac{1}{\left(\frac{dW}{dx}\right)^{1-\alpha}} \hat{p}_x \frac{1}{\left(\frac{dW}{dx}\right)^{\alpha}} + \frac{1}{\left(\frac{dW}{dx}\right)^{\alpha}} \hat{p}_x \frac{1}{\left(\frac{dW}{dx}\right)^{1-\alpha}}\right] \phi_{p_W}(x) = p_W \phi_{p_W}(x) \ . \quad (32)$$

This is easily seen to give

$$\frac{-i}{\left(\frac{dW}{dx}\right)} \frac{d\phi_{p_W}}{dx} - \frac{i}{2} \phi_{p_W} \frac{d}{dx} \frac{1}{\left(\frac{dW}{dx}\right)} = p_W \phi_{p_W}(x) \ . \quad (33)$$

But

$$\frac{d}{dx}\left(\frac{dW}{dx}\right)^{-1} = -\frac{d^2W}{dx^2}\left(\frac{dW}{dx}\right)^{-2} , \quad (34)$$

**so we obtain the α-independent result**

$$\frac{-i}{\left(\frac{dW}{dx}\right)} \frac{d\phi_{p_W}}{dx} + \frac{i}{2} \phi_{p_W} \frac{d^2W}{dx^2}\left(\frac{dW}{dx}\right)^{-2} = p_W \phi_{p_W}(x) \ . \quad (35)$$

The operator $\frac{-i}{\left(\frac{dW}{dx}\right)} \frac{d}{dx} + \frac{i}{2} \frac{d^2W}{dx^2}\left(\frac{dW}{dx}\right)^{-2}$ is self-adjoint under the measure "$dx$". Equation (35) is readily integrated and yields the final result

$$\phi_{p_W}(x) = \phi_{p_W}(0)\left(\frac{dW}{dx}\right)^{1/2} e^{ip_W W} = \phi_{p_W}(0)\left(\frac{dW}{dx}\right)^{1/2} e^{iW(k)W(x)} \quad . \quad (36)$$

This is identical, of course, to the result, Eq. (30) for the special, self-adjoint choice $\alpha = 1/2$. **This demonstrates that the eigenstates of all of the self-adjoint $\hat{p}_W$ are identical, independent of the choice of α!** It is easily seen that the proper normalization is the usual $\phi_{p_W}(0) = \sqrt{1/2\pi}$.

## VI. Non-unitary Transformations and the New Momentum Operators

We recognize that one can always generate MUBs by unitary transformations of existing MUBs. However, this doesn't lead to new MUBs in the present sense. We stress here our new MUBs are not the result of unitary transformations. This is clear from the fact that we generate non-self adjoint momentum operators. For example, the fact that in Eq. (8), the new momentum operator results from a one-sided application of the operator $\dfrac{1}{\left(\dfrac{dW}{dx}\right)}$ to the x-representation momentum operator is new and opens up interesting possibilities. Indeed, in Eq. (6), we again see that the transformation from the old to the new momentum operator does not correspond to a standard transformation in quantum mechanics. In addition, all the α-dependent momentum operators can be related to that of Eq. (8) via similarity transformations. We define the general similarity transform

$$\hat{S} = \left(\frac{dW}{dx}\right)^{-\alpha}, \quad \hat{S}^{-1} = \left(\frac{dW}{dx}\right)^{\alpha}. \quad (37)$$

Then it is easily seen that

$$\frac{-i}{\left(\frac{dW}{dx}\right)}\frac{d}{dx} = \hat{S} \frac{-i}{\left(\frac{dW}{dx}\right)^{1-\alpha}} \frac{d}{dx} \frac{1}{\left(\frac{dW}{dx}\right)^{\alpha}} \hat{S}^{-1} . \quad (38)$$

Thus, all the possible expressions, Eq. (6) (and Eq. (7), since by Eq. (35), it is independent of α), are equivalent, under a similarity transformation, to Eq. (8), which is self-adjoint under the measure $dW = dx\left(\frac{dW}{dx}\right)$. The measure is automatically generated by the canonical transformation. We also point out that one could have chosen the constant of integration, $g(x)$ in Eq. (6) to ensure that one obtains Eq. (8), independent of the choice of $\alpha$.

## VII. Generalized Harmonic Oscillators, Generalized Coherent States and Generalized Fourier Transforms

In this Section, we point out that the operators $\hat{W}$, $\hat{p}_W$ can also be used to define a generalized harmonic oscillator with Hamiltonian

$$\hat{H}_W = \frac{1}{2}\left(\hat{p}_W^2 + \hat{W}^2\right), \quad (39)$$

with $\hat{W}$, $\hat{p}_W$ given by Eq. (11). The eigenstates are given by

$$\psi_j(W) = c_j H_j(W) \exp[-\frac{W^2}{2}] , \quad (40)$$

where the coefficients $c_j$ are the usual normalization constants and the functions $H_j(W)$ are standard Hermite polynomials of the variable W. They are a complete, orthonormal basis, so that

$$\sum_{j=0}^{\infty} \psi^*_j(W) \psi_j(W') = \delta(W-W') \quad (41)$$

and

$$\int_{-\infty}^{\infty} dW \psi^*_j(W') \psi_{j'}(W') = \delta_{jj'} . \quad (42)$$

Just as in the case of the standard harmonic oscillator, where $\Delta x \Delta k = 1/2$ for the ground state, the W-harmonic oscillator ground state for a given polynomial W(x) will satisfy $\Delta W(x) \Delta W(k) = 1/2$. Next we note that Eq. (39) can be factored in terms of the ladder operators

$$\hat{a}_W = \frac{1}{\sqrt{2}} \left( \frac{d}{dW} + W \right) \quad (43)$$

$$\hat{a}_W^+ = \frac{1}{\sqrt{2}} \left( -\frac{d}{dW} + W \right). \quad (44)$$

Thus,

$$\hat{H}_W = \hat{a}_W^+ \hat{a}_W + 1/2 \quad (45)$$

$$= \hat{a}_W \hat{a}_W^+ - 1/2. \quad (46)$$

It follows that one can now construct coherent states using the W-displacement operator or as eigenvectors of the lowering operator. The resulting coherent states are over-complete and can be used as basis-functions for a variety of calculations. They also can be used for new types of semiclassical approximations. In addition, one can generate W-generalized Wigner distributions in the $W, P_W$ phase space. Again, as for the new MUB, these all may be implemented in the coordinate representation with the correct measure guaranteed by the canonical transformation.

We also point out that the orthonormal eigenstates of the W-harmonic oscillator are eigenstates of the W-momentum eigenstates (interpreted as generalized Fourier transform kernels). That is, the $\psi_j(W')$ satisfy the generalized Fourier transform relation

$$\left\langle \phi_{p_W} \Big| \int_{-\infty}^{\infty} dW \, |W\rangle\langle W| \psi_j \right\rangle = \frac{1}{\sqrt{2\pi}} \int_{-\infty}^{\infty} dW \, e^{-ip_W W} \psi_j(W) = \psi_j(p_W). \quad (47)$$

In the coordinate representation, this is

$$\frac{1}{\sqrt{2\pi}} \int_{-\infty}^{\infty} dx \, \frac{dW}{dx} e^{-iW(k)W(x)} \psi_j(W(x)) = \psi_j(W(k)), \quad (48)$$

and it is in this form that the transformation will be used. We also note that time-frequency analyses can be carried out by a windowed generalized Fourier transform, with the window being the ground state of the W-harmonic oscillator, or any other convenient window. Such transformations should be well suited for analysis of chirps.

Finally, we note that one can also use the W-Gaussian to generate new "minimum uncertainty wavelets" and the closely related "Hermite Distributed Approximating Functionals" that have proved to be extremely useful computational tools in a number of areas, as well as for digital signal processing [36-42].

**VIII. Discussion and Conclusions**

First, we stress that for canonical transformations to Cartesian-like variables, Dirac quantization results in the unique operators $\hat{W} = W$, $\hat{p}_W = -i\frac{d}{dW}$. Additionally, simple replacement of the usual position and momentum variables by $\hat{x} = x$, $\hat{p} = -i\frac{d}{dx}$ in any of the infinitely many classical expressions for $W$, $p_W$ typically leads to non-Hermitian operators. These are normally rejected as valid operators but they can all be transformed to the W-representation Hermitian operators, $\hat{W}$, $\hat{p}_W$. The non-Hermitian x-representation operators yield biorthogonal, complete basis sets and the Hermitian cases yield a unique orthonormal complete basis set (all with Dirac delta normalization). These non-Hermitian operators all are examples of quasi-Hermitian operators.

Second, we discuss the conditions under which we can obtain MUBs in the coordinate representation. We've shown that possessing the canonical commutation relation corresponding to a unit Poisson bracket is not sufficient [23]. The standard method of defining a self-adjoint momentum operator as the average of the corresponding classical expressions satisfies the correct commutation relation but does not lead directly to MUBs. Neither does the symmetric or asymmetric quantization of the canonical momentum as defined in Eq. (6), directly yield an MUB in the coordinate representation. Rather, it is only in the new, W-representation that one obtains an MUB [23]. In addition, for any given choice of W, this MUB is unique (as seen when expressed in the x-representation) and independent of the particular way in which one arranges the canonical momentum prior to quantization!

Third, we restricted ourselves to odd power dominated, non-negative coefficient polynomial choices of the generalized position. We ask now what happens if we choose only the even powered polynomials. In that case, the eigenstates of $\hat{p}_W$ have exactly the same form as before. However, there are now additional, standard, normalizable eigenstates in the $L^2$ sense for complex eigenvalues
$p_W = p_W^{real} + i\, p_W^{imag} = W^{real}(k) + iW^{imag}(k)$, of the form

$$\varphi_W(x) = \exp[iW^{real}(k)W(x) - W^{imag}(k)W(x)]\,, \quad (49)$$

$$W(x) \xrightarrow{|x|\to\infty} +\infty\,,\ W^{imag}(k) > 0. \quad (50)$$

These tend to zero as $x \to \pm\infty$ and so the eigenstate structure is more complicated. In addition, their modulus is not constant and eigenstates for different eigenvalues are not orthogonal. A third important issue is the fact that the domain of W with even powers of x is $0 \leq W < \infty$. Thus, the new variable is not "Cartesian-like", since one no longer has the full real line as the domain of the new variable but only the half line. The transform is also no longer invertible **except on the half line,** $0 \leq x < \infty$. This is suggestive of a "radial-like" behavior, and is a situation which will be studied further, along with more general choices of W (e.g., fractional powers of $x$ and others).

Fourth, we see that it is possible to obtain new, unexpected self-adjoint operators by properly accounting for the Jacobian of the canonical transformation. This avoids forcing self-adjointness by symetrization techniques by taking advantage of the fact that the chain rule automatically ensures that one particular ordering of the classical variables, Eq. (8), is manifestly self-adjoint, with the proper measure. There is a sense in which this is an obvious point (it is well known that there are operators that are self-adjoint under one measure and not another). The point is that the change in measure is a natural consequence of our seeking to find more natural coordinates to describe the systems of interest. We stress that this has already lead us in previous studies to develop new coherent states whose convergence properties for excited states are superior to bases that are not defined using information about the ground state [31]. Our strategy here is simply to use the same ideas that make canonical transformations so useful in classical dynamics for the quantum mechanical case. The result is that we now have an infinite number of W choices resulting in sets of operators whose eigenstates form continuous MUBs, as well as continuous, complete biorthogonal and orthogonal basis sets and over-complete coherent states!

Fifth, we have succeeded in constructing 4 distinct types of bases. These include the MUB generated by the three operators, $\{\hat{W}, \hat{p}_W, \hat{W} + \hat{p}_W\}$. Because these can be implemented in the x-representation (with the new measure automatically taken into account), these new MUB make possible an added layer of security in applications to quantum cryptography. Next we have used the W-representation "position" and "momentum" to define the Hamiltonian of a generalized harmonic oscillator, resulting in the W-representation orthonormal basis functions. Again, these will typically be employed in the x-representation. We have generated new continuous, complete biorthogonal and orthogonal bases using the x-representation of the various $\hat{W}$, $\hat{p}_W$ operators. This is despite the fact that the relevant operators are not self adjoint in the x-representation. We also have used the fact that the W-representation harmonic oscillator Hamiltonian can be factored into the corresponding raising and lowering operators to generate new coherent states. Additionally, we also obtain new generalized Wigner distributions based on the W-HO ground states. We note that the eigenstates of $\hat{p}_W$ (in the x-representation) generalize the Fourier transform, so that we have new tools (including new windowed, non-linear transforms) to carry out signal processing of non-linear, non-stationary time-frequency signals (e.g., chirps) that are not amenable to the standard Fourier transform. The eigenstates of $\hat{W} + \hat{p}_W$ (Eq. (14)) are clearly linear chirps in the variable W (and therefore highly non-linear chirps when expressed as functions of x).

Sixth, it is useful to illustrate the robustness of one realization of the MUB in terms of a sparseness of representation condition. The prime example against which we compare is the continuous MUB arising from the operator set $\{\hat{x}, \hat{p}_x, \hat{x} + \hat{p}_x\}$. In this case, we deal with the Fourier transform and recognize that the Gaussian, $\exp[-x^2/2]$, is invariant under it. Other functions that are narrower than the Gaussian (e.g., $\exp[-x^6/2]$, corresponding (roughly) to the SUSY definition $W(x) = x^5$ or precisely to the W-harmonic oscillator choice of $W(x) = x^3$) in the x-representation have much slower decay than the Gaussian after Fourier transforming to the k-representation. This means that their Fourier k-domain representation is not sparse. Thus, the Fourier transform is the optimum basis for representing the Gaussian; it gives the sparsest representation possible for that function but not for others. In like manner, the W-harmonic oscillator choice of $W = x^3$ leads to $\exp[-x^6/2]$, which is invariant under the generalized Fourier transform kernel $\phi_{p_W} = \frac{e^{-ik^3 x^3}}{\sqrt{2\pi}}$. Thus, as expected, we have obtained the optimum (sparsest) basis for describing a system having the ground state $\exp[-x^6/2]$. In addition, we also expect this generalized Fourier transform will be well suited for describing other states that are characterized by the same "coordinate" W(x) [31].

Seventh, we have shown how canonically conjugate transformations lead automatically to quasi-Hermitian operators, such that no symmetrization is required to obtain the valid operator observables, $\hat{W}$, $\hat{p}_W$ and $\hat{W} + \hat{p}_W$. The Jacobian of the canonical transformation automatically supplies the required measure or metric for self-adjointness. This also means that the new complete sets can be implemented in the coordinate representation because the proper measure or metric is known.

In conclusion, we have shown that for canonical transformations between variables which have Cartesian domains, Dirac canonical quantization ensures that the resulting $\hat{W}$, $\hat{p}_W$ is unique and self adjoint in the W-representation. We remark that this is also true for the $p_W$-representation, with operators $p_W, -i\dfrac{d}{dp_W}$. We also have constructed an infinity of new operators generating continuous MUBs, as well as biorthogonal or orthogonal bases. These provide an infinite variety of new MUB sets that can be used for quantum encryption [23]. In addition, they lead to an infinite variety of new, complete bases which can be used as computational tools and for the analysis of signals [36-42].

**ACKNOWLEDGEMENT**

This research was supported in part under R. A. Welch Foundation Grant E-0608. The author D. J. K. also gratefully acknowledges partial support of this research from The Fritz Haber Research Center for Molecular Dynamics of the Hebrew University of Jerusalem from 6/16 through 7/16, under the auspices of Professor R. Baer. Extensive discussions about Mutually Unbiased Bases with H. S. Eisenberg are gratefully acknowledged. Several helpful comments on the manuscript by J. Klauder are also gratefully acknowledged.

<div style="text-align:center">**REFERENCES**</div>